\documentclass[conference,10pt]{IEEEtran}

\usepackage{cite}

\usepackage{graphics,amssymb,amsmath,epsfig,color,textgreek}
\usepackage{graphicx}

\usepackage[dvipsnames]{xcolor}
\usepackage{dcolumn}
\usepackage{bm}
\usepackage{hyperref}
\hypersetup{colorlinks=true,citecolor=cyan,linkcolor=red,urlcolor=magenta}
\usepackage{braket}

\usepackage[caption=false]{subfig}
\usepackage{qcircuit}

\usepackage[dvipsnames]{xcolor}
\usepackage{multirow}

\begin{document}
\title{Efficient Classical Processing of Constant-Depth Time Evolution Circuits in Control Hardware}

\author{
    \IEEEauthorblockN{Akhil Francis\IEEEauthorrefmark{1}\textsuperscript{\textsection},
    Abhi D.~Rajagopala\IEEEauthorrefmark{1}\textsuperscript{\textsection},
    Norm M. Tubman\IEEEauthorrefmark{2},
    Katherine Klymko\IEEEauthorrefmark{3}, and
    Kasra Nowrouzi\IEEEauthorrefmark{1}
    }\\
    \IEEEauthorblockA{
    \IEEEauthorrefmark{1}\textit{AMCR, Lawrence Berkeley National Laboratory, Berkeley, CA, USA}\\
    \IEEEauthorrefmark{2}\textit{NASA Ames Research Center, CA, USA}\\
    \IEEEauthorrefmark{3}\textit{NERSC, Lawrence Berkeley National Laboratory, Berkeley, CA, USA}\\
    }
}

\maketitle
\begingroup\renewcommand\thefootnote{\textsection}
\begin{NoHyper}
\footnotetext{Equal contribution.}
\end{NoHyper}
\endgroup

\begin{abstract}
 
Improving quantum algorithms run-time performance involves several strategies such as reducing the quantum gate counts, decreasing the number of measurements, advancement in QPU technology for faster gate operations, or optimizing the classical processing. This work focuses on the latter, specifically reducing classical processing and compilation time via hardware-assisted parameterized circuit execution (PCE) for computing dynamical properties of quantum systems. PCE was previously validated for QCVV protocols, which leverages structural circuit equivalencies. We demonstrate the applicability of this approach to computing dynamical properties of quantum many-body systems using structurally equivalent time evolution circuits, specifically calculating correlation functions of spin models using constant-depth circuits generated via Cartan decomposition. Implementing this for spin-spin correlation functions in Transverse field XY (up to 6-sites) and Heisenberg spin models (up to 3-sites), we observed a run-time reduction of up to 50\% compared to standard compilation methods. This highlights the adaptability of time-evolution circuit with hardware-assisted PCE to potentially mitigate the classical bottlenecks in near-term quantum algorithms.

\end{abstract}

\section{Introduction}
Computing dynamical properties is an important aspect of studying quantum many-body systems. For example, Green's functions \cite{gomes2023computing} and associated spectral functions \cite{endo2020calculation} are often computed when studying strongly correlated systems. 
Dynamical properties are also relevant in various spectroscopy methods, such as noise spectroscopy \cite{vezvaee2024fourier}, and in calculating transport or diffusion related properties such as Anderson localization \cite{anderson1958absence, kokcu2022fixed}.
Thus, having efficient methods to compute these dynamical quantities is important.  Although there are various classical methods for performing these calculations, such as Exact Diagonalization (ED) or Density Matrix Renormalization Group (DMRG), these classical methods have limitations when applied to large interacting models. Quantum computers offer the potential to simulate and compute dynamical properties for systems that are classically intractable.

Correlation functions are an important dynamical property \cite{bartschi2024potential} to compute Greens functions and spectral functions \cite{gomes2023computing,endo2020calculation,bauer2016hybrid,greene2023quantum }, to determine dynamical structure factors (often probed using inelastic neutron scattering experiments\cite{chiesa2019quantum, baez2020dynamical}), and to study random quantum systems \cite{jain2025dynamical}. Quantum methods to compute correlation functions include the Hadamard test \cite{somma2002simulating, pedernales2014efficient, chiesa2019quantum, francis2020quantum} or related methods \cite{del2024robust }, linear response approaches \cite{kokcu2024linear}, variational approaches \cite{gomes2023computing} or  subspace methods \cite{jamet2205quantum}. These methods require implementation of time evolution operators to simulate the dynamics of the system. 
This could be performed using various approaches such as using Trotter decomposition, which is noisy intermediate-scale quantum (NISQ) friendly and straightforward to implement, or using fault-tolerant suited algorithms such as quantum signal processing \cite{low2017optimal} or Qubitization. 
Another approach is to use constant-depth methods such as variational fast forwarding (VFF) \cite{cirstoiu2020variational} or Cartan decomposition based methods \cite{kokcu2022fixed}, which have been also applied to find correlation functions~\cite{steckmann2023mapping, wan2024hybrid, eassa2023high}.

A key feature of these constant-depth time evolution circuits is that evolving the system requires updating parameters in single-qubit gates for different time steps. This process generates batches of structurally equivalent circuits that differ only in their parameter values. Executing these circuits traditionally requires compiling each one individually into a native hardware format (often representing analog control signals), a step that consumes significant classical compilation time. To address this inefficiency, a hardware-software co-design method, \emph{hardware-assisted parameterized circuit execution (PCE)} \cite{rajagopala2024hardware}, was recently introduced to improve execution efficiency for such circuits. The PCE software automatically identifies structural similarities, creates a reusable circuit template for the batch, and extracts the varying parameters. Complementary hardware design on an FPGA then efficiently recreates the specific circuits using the template and runtime parameter information. This approach substantially reduces compilation time, as only a template of circuits needs initial compilation, and the recreation on FPGA minimizes classical overhead. The method's effectiveness was demonstrated by achieving significant speedups for tomography and benchmarking protocols.

In this work, we apply the hardware-assisted PCE co-design approach to accelerate the computation of dynamical properties by combining it with constant-depth time evolution. Specifically, we compute correlation functions for TFXY and Heisenberg spin models using the Hadamard test, implementing the required time evolution with constant-depth circuits generated via Cartan decomposition. These circuits are then processed using the hardware-assisted PCE framework, and we measure the execution runtime on the \texttt{QubiC}~\cite{xu2023qubic} quantum control system at Advanced Quantum Testbed (AQT) at Berkeley Lab. The remainder of this paper is organized as follows: Section \ref{Methods} describes our methodology. Section \ref{Results} presents the results of applying our approach to the Transverse Field XY (TFXY) and Heisenberg spin models. Finally, Section \ref{Discussion} discusses our findings and conclusions.

\section{Methods}
\label{Methods}
Our co-design approach combines existing  theoretical and experimental methods to calculate correlation for models of interest. We briefly describe the theoretical or algorithmic part first -- the computation of correlation functions using Hadamard tests\cite{somma2002simulating, pedernales2014efficient, del2024robust} in subsection \ref{subsection: Correlation}, and in the following subsection \ref{subsection: TE Cartan} the time evolution method using Cartan decomposition, which results in structurally equivalent circuits. In the final subsection \ref{subsection: hardware assisted PCE}, we describe the classical-processing method, the hardware assisted PCE, which makes use of this structural equivalence.

\subsection{Correlation functions using Hadamard test method}
\label{subsection: Correlation}
The Hadamard test \cite{somma2002simulating} has been used to find correlation functions for time evolution circuits using Trotterization \cite{chiesa2019quantum}, constant-depth methods \cite{francis2020quantum, eassa2023high} and Cartan decomposition \cite{steckmann2023mapping}.
Correlation function between operators $A$ at initial time $0$ and $B$ at later time $t$ for a system in state $\ket{\Phi}$ whose interactions are described by a time independent Hamiltonian $H$ is defined as given in Eq.\eqref{eq:C(t)}. 
\begin{equation}
    C(t) = \bra{\Phi}  B (t) A(0)\ket{\Phi} \ .
    \label{eq:C(t)}
\end{equation}
In the Heisenberg representation, $B(t)$ is expanded as $U(t)^{\dagger}BU(t)$ where $U(t) = e^{-iHt}$ is the time evolution operator. 

\begin{figure}[htpb]
    \centering
    \includegraphics[width=0.99\linewidth]{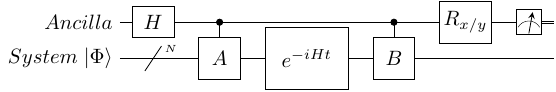}
    \caption{Circuit diagram for Hadamard test method to measure correlation functions given in Eq. \eqref{eq:C(t)}. Here $e^{-iHt}$ denotes the time evolution operator and $R_{x/y}$ denotes basis transformation gates for $X$ and $Y$ basis measurements for the real and imaginary part of the correlation function.}
    \label{fig:Hadamard test}
\end{figure}

This function can be computed using the Hadamard test \cite{pedernales2014efficient}, which uses an ancilla qubit. We briefly outline the method here along with the circuit diagram in Fig.\ref{fig:Hadamard test}. In this method, the ancilla qubit is initialized with a Hadamard gate and the system qubits are prepared in the desired state $\ket{\Phi}$. 
Then a controlled-$A$ operation is applied, followed by time evolution of system qubits, followed by a controled-$B$ operation. Finally the real and imaginary part of the correlation functions can be measured from the expectation values using the appropriate rotation gates ($H$, $R_x(\pi/2)$) to the $X$ and $Y$ basis respectively on the ancilla qubits \cite{pedernales2014efficient, chiesa2019quantum, francis2020quantum}.

Here $A$ and $B$ are assumed to be unitary operators for implementation on quantum hardware. If these are not unitary, they can be written as linear combination of unitaries to employ the method \cite{keen2020quantum}. 
In our case, we study spin models and the spin operators considered are naturally unitary and are expressed using Pauli operators. 
The initial state of the system needs to be known or prepared accordingly. For example, the ground state of the system is studied for many applications \cite{keen2020quantum, chiesa2019quantum, francis2020quantum} which needs an appropriate state preparation method. 
Here, since we are focused on demonstration and profiling, we initialize the state to the $|0\rangle^{\otimes N}$ state for simplicity. Thus, the main non-trivial part in implementing the algorithm is the time evolution, which could be implemented using several approaches. 
For our application, where structural equivalence of circuits is favored, we use constant-depth circuits based on Cartan decomposition to perform the time evolution which is discussed in the next subsection.

\subsection{Time evolution using Cartan decomposition}
\label{subsection: TE Cartan}
The model of interest is described by the respective Hamiltonian, which is typically written as a sum of Hermitian terms such as Pauli terms. 
The dynamics of the system is evaluated by the corresponding time evolution operator, expressed as $U(t) = e^{-iHt}$, for time-independent Hamiltonians. Cartan decomposition-based Hamiltonian simulation is a way to perform time evolution using constant-depth circuits without Trotter error \cite{kokcu2022fixed}.

\begin{figure}[htpb]
    \centering
    \includegraphics[width=0.6\linewidth]{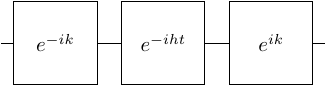}
    \caption{Representation of Cartan based KHK decomposition for the time evolution operator $U(t) = e^{-iHt}$ described in reference \cite{kokcu2022fixed}. Time $t$ is updated in the gate angles corresponding to just the $h$ part. This results in structurally equivalent circuits for different times.}
    \label{fig:KhK evolution}
\end{figure}

Here we briefly outline the method which is described in detail in the reference \cite{kokcu2022fixed}. First, based on the Hamiltonian terms, a full closure algebra ($\mathfrak{g}$) is obtained which is then decomposed into two subsets ($\mathfrak{k}$ and $\mathfrak{m}$), where Hamiltonian elements are within $\mathfrak{m}$, often using an appropriate involution, and further a maximal abelian set ($\mathfrak{h}$) is found within $\mathfrak{m}$. This allows the time evolution operator to appropriately decompose as a product of unitaries (KHK decomposition) assuming no Trotter error as given in Eq.\eqref{eq: Cartan} and represented in Fig.\ref{fig:KhK evolution}.
\begin{equation}
    U(t) = e^{-iHt} = K \prod_{j}e^{-ia_jh^jt} K^\dagger \ ,
    \label{eq: Cartan}
\end{equation}
where $h_j$ are elements in $\mathfrak{h}$ and
K could be further expressed as $K = \prod_{j}e^{ib_jk^j}$ where $k_j$ are elements in $\mathfrak{k}$.
Based on this decomposition, a cost function is minimized using classical optimization to get the appropriate coefficients $a_j$ and $b_j$. 
Once these coefficients are found, the time evolution operator is fully expressed as a product formula, using exponential of Pauli terms, where time $t$ is just a variable. 

The exponentials of Pauli operators can be implemented through Pauli gadgets that use CNOT and single qubit gates \cite{kokcu2022fixed}. 
Thus, once the Cartan decomposition and the corresponding coefficients are determined, then the time evolution for any time $t$ can be employed by just changing the appropriate rotation angles of the single qubit gates in the Pauli gadgets (and only for the set corresponding to $\mathfrak{h}$).

\subsection{Hardware-assisted parameterized circuit execution}
\label{subsection: hardware assisted PCE}
Executing time evolution circuits generated via Cartan decomposition typically requires compiling a separate circuit for each time step \emph{t}. This approach is inefficient, leading to significant classical compilation overhead, especially for long evolution times (\emph{t}) or large system sizes. To address this, prior work by Rajagopala \texttt{et al.}~\cite{rajagopala2024hardware} introduced the hardware-assisted parameterized circuit (PCE) execution framework, which improves execution efficiency by exploiting structural similarities within batches of circuits. To test the time evolution circuits, we use the DOE's Advanced Quantum Testbed (AQT) with QubiC control system running a hardware-assisted PCE design. The system setup consists of a host computer running desktop processor connected to the control system via Ethernet. 

The time evolution circuits are initially generated using the \emph{Qiskit} framework~\cite{qiskit2024}. These circuits are then transpiled using the Qiskit compiler, constrained to a basic gate set (H, CX, CY, CZ, RX, RY, RZ) and a QPU ring topology. This specific transpilation avoids aggressive compiler optimizations that could break the required circuit structure and ensures compatibility with the downstream QubiC compiler. The ring topology reflects the connectivity of the superconducting QPU at the AQT. The hardware-assisted PCE framework employs a front-end software running on the host computer, \emph{Read-Identify-Peel (RIP)}, for identifying structural similarities. The \emph{pre-compile stage} of RIP translates the transpiled Qiskit circuits into the OpenQASM 3.0 format. Subsequently, these circuits are converted into the QubiC native gate format. The single-qubit gates within this format, which represent arbitrary rotations, are decomposed into the ZXZXZ unitary structure specified in Eq.~\eqref{eq:zxzxz}. This decomposition parameterizes each single-qubit gate using three phase angles and fixed $X_{\pi/2}$ pulses. The $X_{\pi/2}$  pulses and the native two-qubit gates are then mapped directly to analog pulses via the QubiC intermediate representation. Using the decomposed circuit representation (Eq.~\eqref{eq:zxzxz}), RIP compares circuits within a batch to identify structural similarities. For the constant-depth time evolution circuits considered here, all circuits in a batch share the same structure. Consequently, RIP generates a single circuit template representing this common structure and extracts the varying phase parameters from the single-qubit gates for each individual circuit. The circuit template follows the control system compilation. The compiled circuit template and the extracted parameter lists are binarized for transmission to the QubiC control system.
\begin{equation}\label{eq:zxzxz}
    U_3(\phi, \theta, \lambda) = Z_{\phi - \pi/2} X_{\pi/2} Z_{\pi - \theta} X_{\pi/2} Z_{\lambda - \pi/2} ~.
\end{equation}

The QubiC control system utilizes an AMD ZCU216 Radio Frequency System-on-Chip (RFSoC)~\cite{zcu216}, which integrates ARM processors and an FPGA. A scheduler part of hardware-assisted PCE executes on the ARM core managing the upload of the binarized circuit template and the parameter lists for the batch. Within the FPGA, the hardware-assisted PCE \emph{Stitch} module reconstructs each specific time evolution circuit in real-time by combining the common template with the corresponding parameter set. The module interacts with the distributed processor to generate analog waveform required for the QPU. This approach significantly reduces compilation and data load times compared to sending full circuits individually. The overhead associated with the Stitch module is minimal, as inserting each parameter into the template takes approximately 4 nanoseconds on the FPGA. Furthermore, this PCE technique inherently scales well. Since, longer evolution times (\emph{t}) requires a single circuit template compilation, the potential speedup increases over time \emph{t}.

\section{Results}
\label{Results}
To investigate the approach we focus on one dimensional spin models. 
We compute spin-spin correlation functions using the Hadamard test with using Cartan decomposition. We profile these with and without using hardware-assisted PCE. To implement the Cartan decomposition and time evolution operators we used the code provided in the reference \cite{kokcu2022fixed}. We further used Qiskit for quantum simulator implementations \cite{qiskit2024}. Our hardware has CZ gate as the native gates instead of CNOT gates, thus we further converted CNOT gates to CZ gates while implementing and profiling.

First we consider the Transverse Field XY (TFXY) model with free or random coefficients and open boundary condition \cite{kokcu2022fixed}. The model Hamiltonian for an $N$-site system is given in Eq.\eqref{eq: TFXY ham}, with the nearest neighbor interactions along the XY plane and magnetic field along the $z$ direction. 

\begin{equation}
    H = \sum_{i=1}^{N-1}J^x_{i}X_iX_{i+1} + J^y_{i}Y_iY_{i+1} + \sum_{i=1}^{N}b_i Z_i \ .
    \label{eq: TFXY ham}
\end{equation}

For simplicity, we compute correlation functions for the state $|0\rangle^{\otimes N}$, which is easy to prepare and we measure the on-site $\langle X_1X_1(t)\rangle$ correlation functions.
We profile these functions with and without hardware-assisted PCE, for two to six site systems for TFXY and two and three site systems for Heisenberg for 50 and 500 time points, or circuits for 1000 shots per time point. We used the similar experimental setup described in the original PCE paper~\cite{rajagopala2024hardware} without connecting to the cryogenic QPU. The setup consists of a desktop computer for compilation and a control hardware running hardware-assisted PCE. 

Next we consider the Heisenberg model given in Eq.\eqref{eq: Heisen ham} in the ferromagnetic regime by fixing $J=-1$
\begin{equation}
    H = \sum_{i=1}^{N-1}J (X_iX_{i+1} +Y_iY_{i+1} + Z_iZ_{i+1} ) \ .
    \label{eq: Heisen ham}
\end{equation}

\begin{figure}[htpb]
    \centering
    \includegraphics[width = 0.49\textwidth]{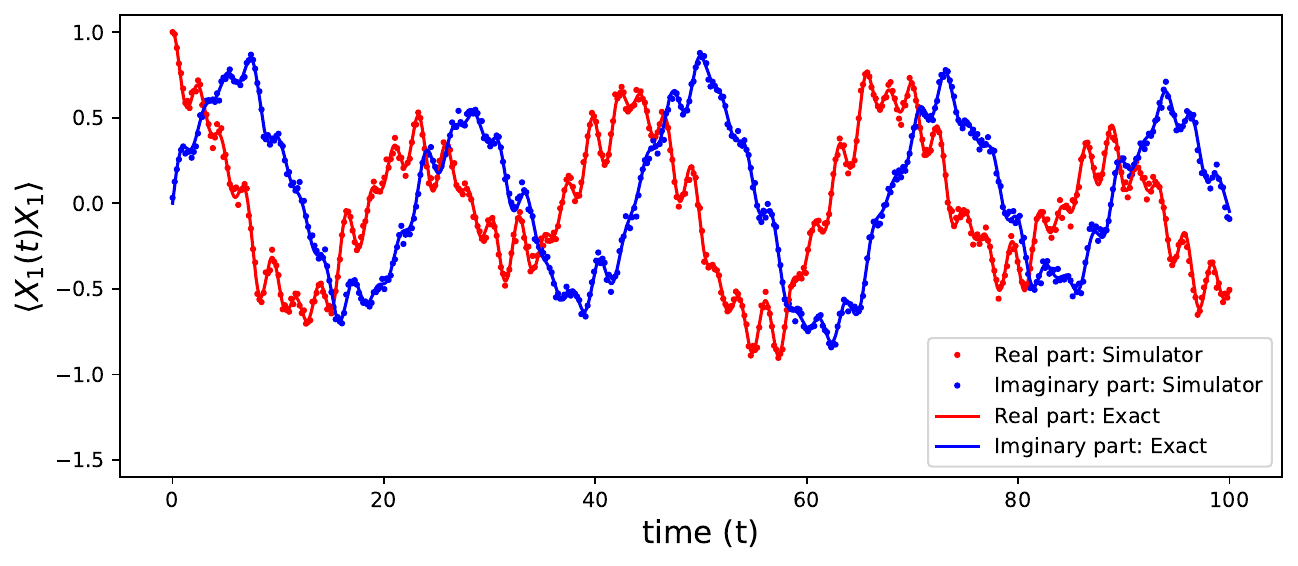}
    \includegraphics[width = 0.49\textwidth]{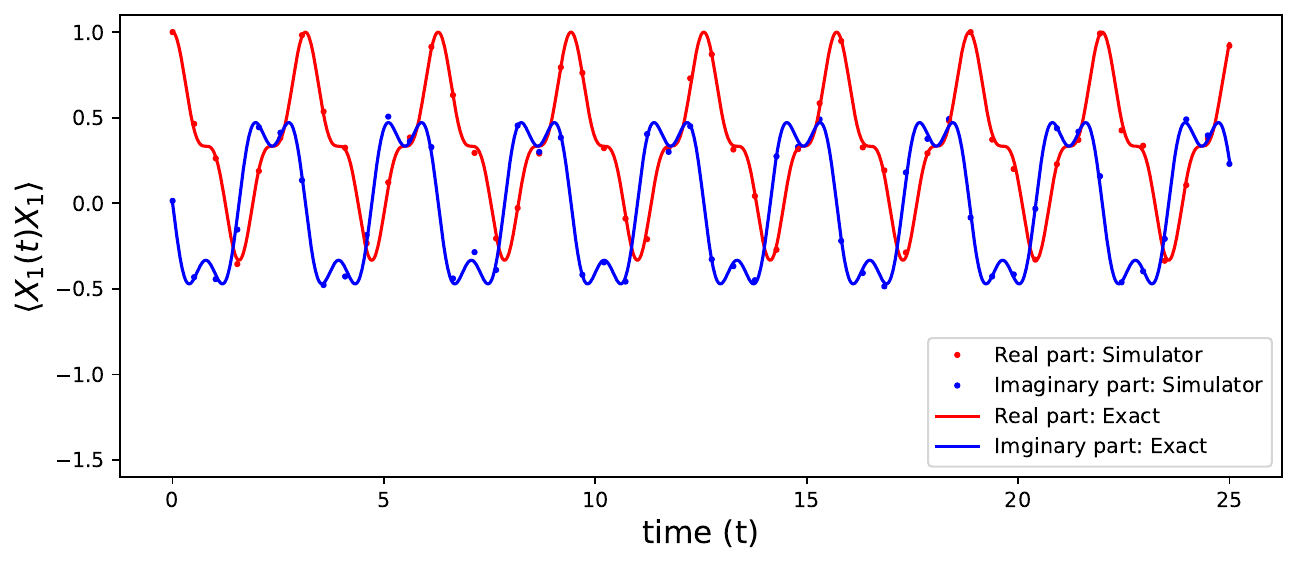}
    \caption{On-site correlation function plots computed for some of the model systems. The real part is given in red color and imaginary part in blue color. Top plot shows correlation functions for six-site TFXY model with random coefficients for 500 time points for the quantum simulator data (shown using solid circles) along with the data from exact diagonalization (shown using solid lines) and bottom plot shows the data for three-site Heisenberg model for 50 time points for the simulator data. For the quantum simulator results we used 1000 shots for each time point.}
    \label{fig: corr fn TFIM plots}
\end{figure}

This model is harder to solve and results in a longer time evolution circuit when compared to the previous model \cite{kokcu2022fixed}. Similarly to the previous model, we compute the $\langle X_1X_1(t)\rangle$ correlation functions of the $|0\rangle^{\otimes N}$ state, for the two and three sites for profiling. Example plots of correlation functions of both the models are shown in Fig.\ref{fig: corr fn TFIM plots}.

The results demonstrate a significant improvement in classical processing performance when using the hardware-assisted PCE framework compared to conventional software-based compilation and execution, as detailed in Table~\ref{tab:Profiledata}. We perform profiling using the profiling infrastructure provided by hardware-assisted PCE. We quantified this improvement by analyzing two metrics: compilation time speedup and total classical time reduction. The compilation speedup is defined as the ratio of compilation time without PCE to that with hardware-assisted PCE. This speedup occurs because hardware-assisted PCE compiles only a single circuit template per batch, whereas the conventional method compiles each circuit individually and converts them into pulse information. As shown in Table~\ref{tab:Profiledata}, for batches corresponding to 50 circuits (time points), the average compilation speedup was $30.59\times$ for the TFXY model (averaged over 2 to 6 sites) and $33.25\times$ for the Heisenberg model (averaged for 2 and 3 sites). For larger batches to 500 circuits (time points), the average compilation speedup increased substantially to $251.07\times$ (TFXY model, 2 to 6 sites) and $260.12\times$ (Heisenberg 2 and 3 sites). We also evaluated the reduction in total classical time (defined as total execution time minus quantum runtime). Comparing scenarios with and without PCE, the hardware-assisted framework reduced the total classical processing time by an average of approximately $50\%$ across the tested configurations reported in Table~\ref{tab:Profiledata}.

\begin{table}[t]
    \centering
        \caption{Profiling results of TFXY and Heisenberg models using hardware-assisted PCE. The speedup compares the runtime without PCE and with PCE.}
    \label{tab:Profiledata}
    \begin{tabular}{|c|c|r|r|r|r|}
    \hline
    \multirow{2}{*}{Model}&\multirow{2}{*}{\# Circuits}&\multirow{2}{*}{Sites}&
    \multicolumn{1}{c|}{Compile}
    &\multicolumn{1}{c|}{Time}
    &\multicolumn{1}{c|}{Time}\\
    &&&Speedup&Reduced(\%)&Saved (s)\\
    \hline
    \multirow{10}{*}{TFXY}&\multirow{5}{*}{50}&2&30.75&44.00&1.65\\
    &&3&35.33&47.94&2.89\\
    &&4&30.98&46.99&3.97\\
    &&5&35.30&51.07&6.46\\
    &&6&20.59&52.45&8.81\\
    \cline{2-6}
    &\multirow{5}{*}{500}&2&239.81&42.37&13.94\\
    &&3&254.88&46.76&25.32\\
    &&4&257.48&51.48&43.19\\
    &&5&256.70&52.72&62.63\\
    &&6&246.49&53.90&85.84\\
    \hline
    \multirow{4}{*}{Heisenberg}&\multirow{2}{*}{50}&2&32.69&42.48&1.49\\
    &&3&33.81&53.36&4.18\\
    \cline{2-6}
    &\multirow{2}{*}{500}&2&266.33&42.20&13.34\\
    &&3&253.92&56.23&42.25\\
    \hline
    \end{tabular}
\end{table}

\section{Discussion}
\label{Discussion}
In this work, we have shown that constant-depth time evolution circuits can be efficiently compiled and classically processed using hardware-assisted parameterized circuit execution (PCE) to compute dynamical properties such as correlation functions. 
We have demonstrated this approach for the TFXY model with random coefficients and Heisenberg models for smaller system sizes. From the profiled data from hardware-assisted PCE, we see that there is a signification improvement in the compilation and $50\%$ reduction in classical processing time for the circuits with different sites and time points. These results provides the intuition into the time savings which increases for larger site model and longer time points.

We note that even though there is a relevant speed up, the time savings is not significant for the examples considered here. 
This is because the models used are smaller and the overall classical time for these small circuits is not very long. However, this could increase for larger systems and longer evolution time, for applications such as finding dynamical structure factors \cite{bartschi2024potential} or when used as a part of other iterative algorithms such as DMFT \cite{bauer2016hybrid, steckmann2023mapping, bartschi2024potential}.

Other than computing correlation functions, time evolution circuits could also be used to study various other dynamical properties such as transport properties or localization effects \cite{kokcu2022fixed}.

As long as we have structurally similar circuits, which arises when using constant-depth methods such as Cartan decomposition or VFF \cite{cirstoiu2020variational}, the method and experiment described in this work are relevant and the hardware-assisted PCE could be applied for efficient classical processing. 
 
Other than using time evolution and studying dynamical properties, structurally equivalent circuits arise in other applications and algorithms. An important such application is variational algorithms for finding ground states such as Variational Quantum Eigensolvers, where variational parameters are often manifested as changes in the angles of gates, and usually proceeds iteratively due to a classical optimization step. 
We note that efficient classical processing of these algorithms might require parameterizing other parts of the algorithm, such as evaluating and minimizing the cost function. 
We finally comment that it could likely be efficient to combine our approach with error mitigation methods such as randomized compiling that are compatible with hardware-assisted PCE \cite{rajagopala2024hardware}.

Co-design approaches between different stacks of quantum computing would become more relevant in the future as the quantum hardware is improving and more applications are investigated. Here we perform co-design between control system and algorithms part, where we apply hardware-assisted PCE which is a recently developed control system method, for an important application to simulate dynamical properties, in particular, correlation functions. We hope this work would pave the way for more such co-design approaches in the future. 

\section{Acknowledgments}
This work was supported by the Laboratory Directed Research and Development Program of Lawrence Berkeley National Laboratory under U.S. Department of Energy Contract No.~DE-AC02-05CH11231. The hardware-assisted parameterized circuit execution used in this work is protected under the U.S.~patent application no.~PCT/US25/30612, filed by Lawrence Berkeley National Laboratory on behalf of the inventor A.D.R.

\bibliographystyle{ieeetr}
\bibliography{ref}

\end{document}